\documentclass[12pt,prd,tightenlines,nofootinbib,showpacs,showkeys]{revtex4}
\newcommand{\be}{\begin{equation}}
\newcommand{\ee}{\end{equation}}
\newcommand{\bdis}{\begin{displaymath}}
\newcommand{\edis}{\end{displaymath}}
\newcommand{\bga}{\begin{equation}\begin{gathered}}
\newcommand{\ega}{\end{gathered}\end{equation}}
\usepackage{bm}
\usepackage{graphics}
\usepackage{rotating}
\usepackage{epsfig}
\usepackage{amsmath}
\usepackage{amsfonts}

\begin{document}
\title{\begin{flushright}{\rm\normalsize SSU-HEP-13/08\\[5mm]}\end{flushright}
Relativistic corrections to the pair double heavy \\diquark production in $e^+e^-$ annihilation}
\author{\firstname{A.P.} \surname{Martynenko}}
\affiliation{Samara State University, Pavlov Street 1, 443011, Samara, Russia}
\affiliation{Samara State Aerospace University named after S.P. Korolyov, Moskovskoye Shosse 34, 443086,
Samara, Russia}
\author{\firstname{A.M.} \surname{Trunin}}
\affiliation{Samara State Aerospace University named after S.P. Korolyov, Moskovskoye Shosse 34, 443086,
Samara, Russia}

\begin{abstract}
On the basis of perturbative QCD and relativistic quark model we
calculate relativistic and bound state corrections in the production
processes of a pair of double heavy diquarks.
Relativistic factors in the production amplitude connected with the
relative motion of heavy quarks and the transformation law of the
bound state wave function to the reference frame of the moving
diquark $S$-wave bound states are taken into account. For the gluon and quark
propagators entering the production vertex function we use a
truncated expansion in the ratio of the relative quark momenta to
the center-of-mass energy $s$ up to the second order.
Relativistic corrections to the quark-quark bound state wave functions in
the rest frame are considered by means of the Breit-like potential.
It turns out that examined effects change essentially
nonrelativistic results of the cross sections. The estimate of the yield of pairs
of double heavy baryons $(ccq)$ at the B-factory is presented.
\end{abstract}



\maketitle

\section{Introduction}

In last years the reactions of pair charmonium production in $e^+e^-$ annihilation
have attracted considerable interest. A growth of the luminosity made it possible
to observe experimentally at the Belle and BABAR \cite{Belle,BaBar}
double $S$- and $P$-wave charmonium production. On the
other hand, the defects of theoretical description of such processes on the basis
of nonrelativistic QCD (NRQCD) were revealed and corrected
\cite{BL1,bodwin,Chao,Qiao,BLL,EM2006,ji,jia}.
Despite the evident successes achieved in this field on the basis of NRQCD \cite{BBL}
and potential quark models in correcting the discrepancy
between the theory and experiment, the double charmonium production
in $e^+e^-$ annihilation remains an interesting task. On the one
hand, there are other production processes of orbitally excited
charmonium states which can be investigated in the same way as the production
of $S$-wave states. Several years ago the Belle and BABAR collaborations
discovered new charmonium-like states in $e^+e^-$ annihilation \cite{pahlova,brambilla-2011}.
The nature of these numerous resonances remains unclear to the present. Some
of them could be considered as a $D$-wave excitations in
the system ($c\bar c$). On the other hand, the variety of the
used approaches and model parameters in this problem raises a
question about the comparison of obtained results
that will lead to a better understanding of the quark-gluon dynamics and
different mechanisms of double heavy quarkonium production.
At last, the obtained luminosity on the meson B-factory ${\mathcal L}=10^{34}~cm^{-2}~c^{-1}$
allows to observe double heavy baryon
$(ccq)$ production. In the threshold region of double
heavy baryon production in $e^+e^-$ annihilation the double baryon production
can give appreciable contribution to the cross
section. For the estimate of such events yield in \cite{bkc}
it was performed a calculation of exclusive pair production of double heavy diquarks ($\bar{\cal D}$
and ${\cal D}$) in nonrelativistic approximation. It seems reasonably good guess that first stage
of double heavy baryon production in $e^+e^-$ annihilation consists in the formation of the diquark
nuclei $(Q_1Q_2)$ and $(\bar Q_1\bar Q_2)$ which are tightly bound, small size anti-triplet pairs.
\cite{baryon_diquark,gklo}.
In second stage the produced diquark and antidiquark join a light quark to produce the final
baryons $(Q_1Q_2q)$ and $(\bar Q_1\bar Q_2\bar q)$
if we neglect possible formation of ${\cal D}\bar {\cal D}$ bound states. Other baryon production mechanism
in $e^+e^-$ annihilation connected with a production of $Q\bar Q$ pair and its subsequent fragmentation
into the baryons was analyzed also in the literature \cite{brambilla-2011,baryon_diquark}.
So, the first stage of the process looks similar to the double charmonium production. It is clear that for it
theoretical description we can use improved relativistic formalism as in the meson case \cite{EM2006}.

It is useful to remember that two sources of the
change of nonrelativistic cross section for double
charmonium production are revealed to the present: radiative
corrections of order $O(\alpha_s)$ and relative motion of $c$-quarks
forming the bound states. An actual physical processes of charmonium
production require a formation of hadronic particles in final states
(bound states of a charm quark $c$ and a charm antiquark $\bar c$), for
which perturbative quantum chromodynamics can not provide high precision description.
In quark model a transition of free quarks to the mesons is described
in terms of the bound state wave functions.
Further investigation of exclusive heavy quark bound state production in $e^+e^-$ annihilation
including relativistic effects by an example of diquarks
can improve our understanding of a formation of quark bound states.

This work continues our study of the exclusive double
charmonium production in $e^+e^-$ annihilation in the case of a diquark
$(cc)$, $(bc)$ $S$-wave states on the basis of a
relativistic quark model (RQM) \cite{EM2006,EFGM2009,EM2010,apm2005,rqm5,mt}.
Note that the term RQM specifies an approach in which the systematic
account of corrections connected with relative motion of heavy quarks
can be performed. Relativistic quark
model provides a solution in many tasks of heavy quark physics.
It uses a number of perturbative and nonperturbative parameters
entering in the quark interaction operator. All observables can be
expressed in terms of these parameters. In this way we can check the
predictions of any quark model and draw a conclusion about its
successfulness. At the same time the existence of a large number of different
quark models which are sometimes very complicated for the practical use
puts a question about the elaboration of the unified model containing
generally accepted structural elements.
Another approach to the heavy quark physics which does not contain the ambiguities
of the quark models was formulated in \cite{BBL}. As any other model
of strong interactions of quarks and gluons the approach of NRQCD introduces
in the theory a large number of matrix elements parameterizing nonperturbative
dynamics of quarks and gluons \cite{bodwin,BBL,gremm}. To a certain extent the microscopic picture of the
quark-gluon interaction resident in quark models is changed by the global
picture operating with the numerous nonperturbative matrix elements. The
improved determination of color-singlet NRQCD matrix elements for $S$-wave
charmonium is presented in \cite{bodwin}. Their study evidently shows
that the account of relative order $v^2$ corrections significantly increases
the values of the matrix elements of leading order in $v$.
The correspondence between parameters of quark models and NRQCD which can
be established, opens the way for better understanding of quark-gluon interactions
at small distances. In this sense both approaches complement each other and
could reveal new aspects of color dynamics of quarks and gluons. Thus, the aim of
this study consists in the extension of relativistic approach to the quarkonium production
from Refs.\cite{EM2006,EFGM2009,EM2010} on the processes of exclusive pair diquark production
$e^++e^-\to {\mathcal D}+\bar {\mathcal D}$,
investigation the role of relativistic corrections of order $v^2$ to the production amplitudes
and cross sections
and determination of the interrelationship with the predictions of NRQCD. Assuming that arising
in $e^+e^-$ annihilation diquarks can fragment into double heavy baryons we use the obtained expressions
of total cross sections for an estimate of the cross sections for the pair production of baryons.

\section{General formalism}

In the ground state the diquarks are two-particle bound states of quarks in an antisymmetric
color state with zero angular momentum, positive parity and definite flavor and spin.
A diquark may be an axial vector (spin 1) or a scalar (spin 0). In the case of two identical
quarks a diquark has a spin 1. The attractive forces between two quarks in antisymmetric color
state lead to a formation of the bound system which can be described in quark model in a manner
similar to the quark-antiquark states. A diquark constructed from two heavy quarks ($b$ and $c$)
may be considered as a nucleus of double heavy baryon.
The production of heavy quark bound states at different
high energy reactions is an interesting physical process which is studied during many tenths of years
\cite{brambilla-2011,braaten,kramer,brodsky2013}. It gives an opportunity to investigate the quark-gluon dynamics
beginning from small distances where the perturbative QCD is applicable, to large distances where nonperturbative
aspects of QCD become crucial.
We investigate the exclusive diquark-antidiquark  production in electron-positron annihilation in the lowest-order
perturbative quantum chromodynamics.
The final state consists of a pair of bound states $(bc)$ and $(\bar b\bar c)$ with different
spins in the case
of different heavy quarks. The case of two identical quarks $(cc)$ or $(bb)$ leads to the
production only a pair of axial vector diquarks.
The diagrams that give contributions to the amplitude of a diquark pair production processes in
leading order of the QCD coupling constant $\alpha_s$ are presented
in Fig.~\ref{fig:fig1}. Two other diagrams can be obtained by corresponding
permutations. There are two stages of double diquark production process. In the
first stage, which is described by perturbative QCD, the virtual
photon $\gamma^\ast$ and gluon $g^\ast$ produce two heavy quarks $(bc)$ and two heavy antiquarks $(\bar b\bar c)$
with the following four-momenta:

\begin{figure}[t!]
\centering
\includegraphics[width=5.0 cm]{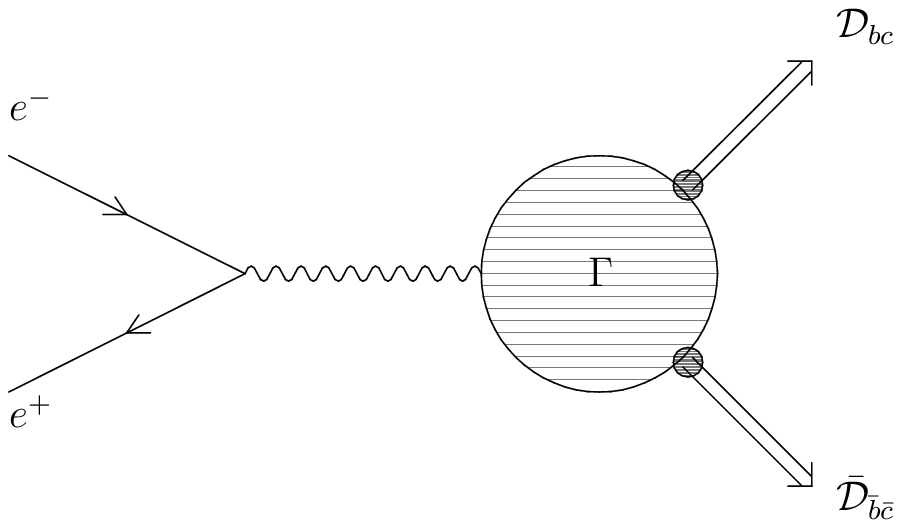}\hspace*{0.4cm}
\includegraphics[width=11.0 cm]{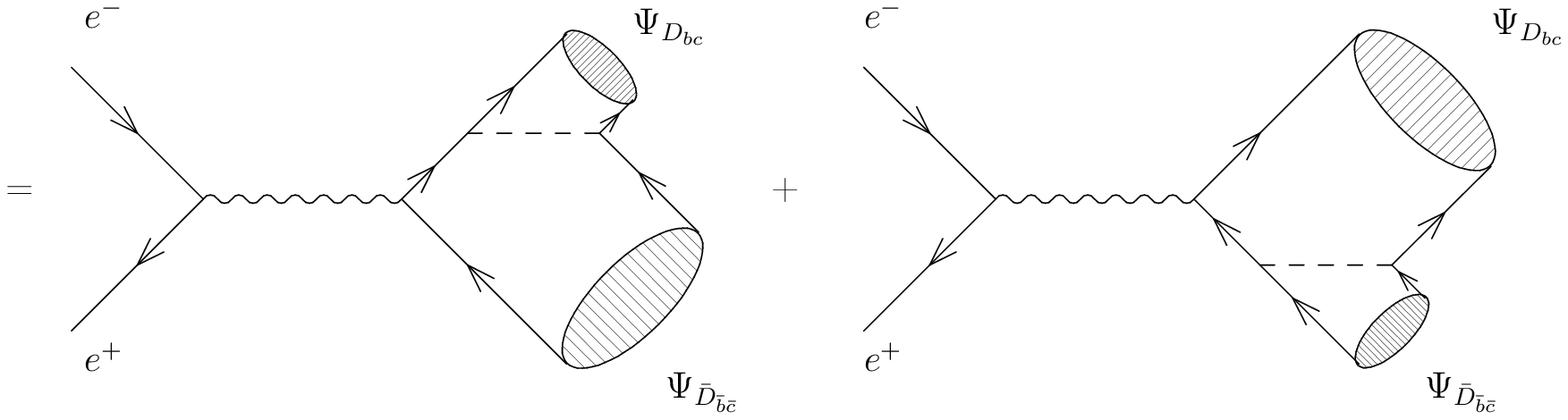}
\caption{The production amplitude of a pair of diquark
states in $e^+e^-$ annihilation. ${\cal D}_{bc}$, $\bar{\cal D}_{\bar b\bar c}$
denote the diquark and antidiquark states composed from heavy quarks $b$ and $c$ and antiquark $\bar b$
and $\bar c$ correspondingly. Wavy line shows
the virtual photon and dashed line corresponds to the gluon.
$\Gamma$ is the production vertex function.}
\label{fig:fig1}
\end{figure}

\begin{equation}
\label{eq:pq}
p_1=\eta_{11}P+p,~p_2=\eta_{12}P-p,~(p\cdot P)=0,~\eta_{1i}=\frac{M_{D_{bc}}^2\pm m_1^2\mp m_2^2}{2M_{D_{bc}}^2},
\end{equation}
\begin{displaymath}
q_1=\eta_{21}Q+q,~q_2=\eta_{22}Q-q,~(q\cdot Q)=0,~\eta_{2i}=\frac{M_{\bar D_{\bar b\bar c}}^2\pm m_1^2\mp m_2^2}{2M_{\bar D_{\bar b\bar c}}^2}
\end{displaymath}
where $M_{D_{bc}}$ is the mass of diquark consisting of quarks $b$ and $c$. $P(Q)$ are the total four-momenta, $p=L_P(0,{\bf p})$,
$q=L_P(0,{\bf q})$ are the relative four-momenta obtained from the
rest frame four-momenta $(0,{\bf p})$ and $(0,{\bf q})$ by the
Lorentz transformation to the system moving with the momenta $P$, $Q$.
The momenta $p_{1,2}$ of the heavy quarks $c$, $b$ and antiquarks $\bar c$, $ \bar b$ are not on
a mass shell: $p_{1,2}^2=\eta_{1i}^2P^2-{\bf p}^2=\eta_{1i}^2M_{D_{bc}}^2-{\bf p}^2\not= m_{1,2}^2$.
An expressions~\eqref{eq:pq}
describe the symmetrical escape of heavy quarks and antiquarks from
the mass shell. In the second nonperturbative stage, quark and antiquark pairs form
double heavy diquarks.

Let consider the production amplitude of scalar and axial vector diquarks, which can be
presented in the form \cite{rqm5,EM2006,EM2010}:
\begin{equation}
\label{eq:amp}
{\cal M}(p_-,p_+,P,Q)=-\frac{8\pi^2\alpha}{3s^2}\sqrt{M_{D_{bc}}M_{\bar D_{\bar b\bar c}}}\bar v(p_+)\gamma^\beta u(p_-)\delta_{ij}
\int\frac{d{\bf p}}{(2\pi)^3}
\int\frac{d{\bf q}}{(2\pi)^3}
\times
\end{equation}
\begin{displaymath}
\times Sp\left\{\bar\Psi^{\cal S}_{D_{bc}}(p,P)\Gamma_1^{\beta\nu}(p,q,P,Q)\bar\Psi^{\cal AV}_{\bar D_{\bar b\bar c}}(q,Q)\gamma_\nu-
\bar\Psi^{\cal S}_{D_{bc}}(-p,P)\Gamma_2^{\beta\nu}(p,q,P,Q)\bar\Psi^{\cal AV}_{\bar D_{\bar b\bar c}}(-q,Q)
\gamma_\nu\right\},
\end{displaymath}
where $s$ is the center-of-mass energy,
a superscript ${\cal S}$ indicates a scalar diquark, a superscript ${\cal AV}$ indicates an axial vector
diquark, $\alpha$ is the fine structure constant. $\Gamma_{1,2}$ are the vertex functions defined below.
The transition of free quarks to diquark bound states is described by specific wave functions.
Relativistic wave functions of scalar and axial vector diquarks accounting for the transformation from the
rest frame to the moving one with four momenta $P$, $Q$ are
\begin{eqnarray}
\label{eq:amp1}
\bar\Psi^{\cal S}_{D_{bc}}(p,P)&=&\frac{\bar\Psi^0_{D_{bc}}({\bf p})}{
\sqrt{\frac{\epsilon_1(p)}{m_1}\frac{(\epsilon_1(p)+m_1)}{2m_1}
\frac{\epsilon_2(p)}{m_2}\frac{(\epsilon_2(p)+m_2)}{2m_2}}}
\left[\frac{\hat v_1-1}{2}+\hat
v_1\frac{{\bf p}^2}{2m_2(\epsilon_2(p)+ m_2)}-\frac{\hat{p}}{2m_2}\right]\cr
&&\times\gamma_5(1+\hat v_1) \left[\frac{\hat
v_1+1}{2}+\hat v_1\frac{{\bf p}^2}{2m_1(\epsilon_1(p)+
m_1)}+\frac{\hat{p}}{2m_1}\right],
\end{eqnarray}
\begin{eqnarray}
\label{eq:amp2}
\bar\Psi^{\cal AV}_{\bar D_{\bar b\bar c}}(q,Q)&=&\frac{\bar\Psi^0_{\bar D_{\bar b\bar c}}({\bf q})}
{\sqrt{\frac{\epsilon_1(q)}{m_1}\frac{(\epsilon_1(q)+m_1)}{2m_1}
\frac{\epsilon_2(q)}{m_2}\frac{(\epsilon_2(q)+m_2)}{2m_2}}}
\left[\frac{\hat v_2-1}{2}+\hat v_2\frac{{\bf q}^2}{2m_1(\epsilon_1(q)+
m_1)}+\frac{\hat{q}}{2m_1}\right]\cr &&\times\hat{\varepsilon}_{\cal
AV}(Q,S_z)(1+\hat v_2) \left[\frac{\hat v_2+1}{2}+\hat
v_2\frac{{\bf q}^2}{2m_2(\epsilon_2(q)+ m_2)}-\frac{\hat{q}}{2m_2}\right],
\end{eqnarray}
where the hat is a notation for the contraction of four vector with
the Dirac matrices, $v_1=P/M_{D_{bc}}$, $v_2=Q/M_{\bar D_{\bar b\bar c}}$;
$\varepsilon_{\cal AV}(Q,S_z)$ is the polarization vector of the
axial vector diquark, $\epsilon_{1,2}(p)=\sqrt{p^2+m_{1,2}^2}$ and $m_{1,2}$
are the masses of $c$ and $b$ quarks. The relativistic functions~\eqref{eq:amp1}-\eqref{eq:amp2} and the vertex functions $\Gamma_{1,2}$
do not contain the $\delta ({\bf p}^2-\eta_{1i}^2M_{D_{bc}}^2+m_{1,2}^2)$.
More complicated factor including the bound state wave function in the rest frame presented
in Eqs.~\eqref{eq:amp1} and \eqref{eq:amp2} plays the role of the $\delta$-function.
This means that instead of the substitutions $M_{D_{bc}}=\epsilon_1({\bf p})+\epsilon_2({\bf p})$ and
$M_{\bar D_{\bar b\bar c}}=\epsilon_1({\bf q})+\epsilon_2({\bf q})$ in the production amplitude we carry out the
integration over the quark relative momenta ${\bf p}$ and ${\bf q}$.
Color part of the diquark wave function in the amplitude~\eqref{eq:amp} is taken as $\epsilon_{ijk}/\sqrt{2}$
(color indexes $i, j, k=1, 2, 3$),
so that general color factor in~\eqref{eq:amp} is equal to $\delta_{ij}$.
Relativistic wave functions in Eqs.~\eqref{eq:amp1} and \eqref{eq:amp2} are equal to the product of wave functions in the rest frame
$\Psi^0_{D_{bc}}$ and spin projection operators that are
accurate at all orders in $|{\bf p}|/m$ \cite{rqm5,EM2006}. An expression of spin projector in different
form has
been derived primarily in \cite{Bodwin2002} where spin projectors are
written in terms of heavy quark momenta $p_{1,2}$ lying on the mass shell.
Our derivation of relations~\eqref{eq:amp1} and \eqref{eq:amp2} accounts for the transformation
law of the bound state wave functions from the rest frame to the
moving one with four momenta $P$ and $Q$. This transformation law
was discussed in the Bethe-Salpeter approach in \cite{BP} and in
quasipotential method in \cite{F1973}. We use the last one and write necessary transformation as follows:
\begin{equation}
\label{eq:trans}
\Psi_{P}^{\rho\omega}({\bf p})=D_1^{1/2,~\rho\alpha}(R^W_{L_{P}})
D_2^{1/2,~\omega\beta}(R^W_{L_{P}})\Psi_{0}^{\alpha\beta}({\bf p}),
\end{equation}
\begin{displaymath}
\bar\Psi_{P}^{\lambda\sigma}({\bf p})
=\bar\Psi^{\varepsilon\tau}_{0}({\bf p})D_1^{+~1/2,~\varepsilon
\lambda}(R^W_{L_{P}})D_2^{+~1/2,~\tau\sigma}(R^W_{L_{P}}),
\end{displaymath}
where $R^W$ is the Wigner rotation, $L_{P}$ is the Lorentz boost
from the diquark rest frame to a moving one, and the rotation matrix
$D^{1/2}(R)$ is defined by
\begin{equation}
{1 \ \ \,0\choose 0 \ \ \,1}D^{1/2}_{1,2}(R^W_{L_{P}})= S^{-1}({\bf
p}_{1,2})S({\bf P})S({\bf p}),
\end{equation}
where explicit form for the Lorentz transformation matrix of four-spinor is
\begin{equation}
S({\bf p})=\sqrt{\frac{\epsilon(p)+m}{2m}}\left(1+\frac{(\bm{\alpha}
{\bf p})} {\epsilon(p)+m}\right).
\end{equation}

We omit here intermediate expressions giving rise to our final relations~\eqref{eq:amp}-\eqref{eq:amp2}
\cite{EM2006,EFGM2009}. The presence of the $\delta (p\cdot P)$ function
allows to make the integration over relative energy $p^0$ if we write the initial
production amplitude as a convolution of the truncated amplitude with two
Bethe-Salpeter (BS) diquark wave functions. In the rest frame of a bound state the condition
$p^0=0$ allows to eliminate the relative energy
from the BS wave function. The BS wave function satisfies a two-body bound state equation
which is very complicated and has no known solution. A way to deal with this problem
is to find a soluble lowest-order equation containing main physical properties
of the exact equation and develop a perturbation theory. For this purpose we continue
to work in three-dimensional quasipotential approach. In this framework the double
diquark production amplitude~\eqref{eq:amp} can be written initially as a product of the production
vertex function $\Gamma_{1,2}$ projected onto the positive energy states by means of the Dirac
bispinors (free quark wave functions) and a bound state quasipotential wave functions
describing diquarks in the reference frames moving with four momenta $P,Q$.
Further transformations include the known transformation law of the bound state wave
functions to the rest frame~\eqref{eq:trans}. The physical
interpretation of the double diquark production amplitude is the following:
we have a complicated transition of two heavy quark and antiquark
which are produced in $e^+e^-$-annihilation outside the mass shell and their
subsequent evolution firstly on the mass shell (free Dirac bispinors) and then to the
quark bound states. In the spin projectors we have
${\bf p}^2\not=\eta_{1i}^2M^2-m_{1,2}^2$ just the same as in the vertex production functions
$\Gamma_{1,2}$.
We can not say exactly whether heavy quarks are on-shell or not in the spin
projectors~\eqref{eq:amp1}-\eqref{eq:amp2} because we should consider these structures as a transition form factors for
heavy quarks from free states to bound states. In the course of the ${\cal M}$ transformation we introduce
symmetrical spin wave functions for vector and scalar diquarks \cite{baryon_diquark,hussain}:
\begin{equation}
u_i(0)u_j(0)=\left[\frac{(1+\gamma_0)}{2\sqrt{2}}\hat\varepsilon_{\cal AV}(\gamma_5)C\right]_{ij},~~~
v_i(0)v_j(0)=\left[\frac{(1-\gamma_0)}{2\sqrt{2}}\hat\varepsilon_{\cal AV}(\gamma_5)C\right]_{ij},
\end{equation}
where $C$ is the charge conjugation matrix. As the color wavefunction of identical quarks $(cc)$ or $(bb)$
is antisymmetric and the quarks are taken to be in the ground state $S$-wave, the spin wave function
must be symmetric. So, the $(cc)$ or $(bb)$ pair can only form a spin 1 diquark.

At leading order in $\alpha_s$ the vertex functions $\Gamma_{1,2}^{\beta\nu}(p,P;q,Q)$ can be written as
($\Gamma_2^{\beta\nu}(p,P;q,Q)$ can be obtained from $\Gamma_1^{\beta\nu}(p,P;q,Q)$ by means of the replacement
$p_1\leftrightarrow p_2$, $q_1\leftrightarrow q_2$, $\alpha_b\to\alpha_c$, $Q_c\to Q_b$)
\begin{equation}
\Gamma_1^{\beta\nu}(p,P;q,Q)= Q_c\alpha_b\left[\gamma_\mu\frac{(\hat l-\hat
q_1+m_1)}{(l-q_1)^2-m_1^2+i\epsilon} \gamma_\beta D^{\mu\nu}(k_2)+
\gamma_\beta\frac{(\hat p_1-\hat l+m_1)}{(p_1-l)^2-m_1^2+i\epsilon}
\gamma_\mu D^{\mu\nu}(k_2)\right],
\end{equation}
\begin{equation}
\Gamma_2^{\beta\nu}(p,P;q,Q)=Q_b\alpha_c\left[\gamma_\mu\frac{(\hat l-\hat q_2+m_2)}{(l-q_2)^2-m_2^2+i\epsilon} \gamma_\beta D^{\mu\nu}(k_1)+
\gamma_\beta\frac{(\hat p_2-\hat l+m_2)}{(p_2-l)^2-m_2^2+i\epsilon}
\gamma_\mu D^{\mu\nu}(k_1)\right],
\end{equation}
where the gluon momenta are $k_1=p_1+q_1$, $k_2=p_2+q_2$ and
$l^2=s^2=(P+Q)^2=(p_-+p_+)^2$, $\alpha_{c,b}=\alpha_s\left(\frac{m_{1,2}^2}{M^2}s^2\right)$,
$p_-$, $p_+$ are four momenta of the
electron and positron. The dependence on the relative momenta of
heavy quarks is presented both in the gluon propagator $D_{\mu\nu}(k)$
and quark propagator as well as in relativistic wave functions~\eqref{eq:amp1} and \eqref{eq:amp2}.
Taking into account that the ratio of relative quark
momenta $p$ and $q$ to the energy $s$ is small, we expand
inverse denominators of quark and gluon propagators as follows:
\begin{equation}
\label{eq:pr1}
\frac{1}{(l-q_{1,2})^2-m_{1,2}^2}=\frac{1}{r_{2,1}s^2}\left[1-\tilde B_{AV}\frac{(2r_{1,2}-1)}{r_{2,1}}
-\frac{2r_{1,2}M^2}{s^2}(\tilde B_S-r_{1,2}\tilde B_{AV})-\frac{(q^2\mp 2lq)}{r_{2,1}s^2}
+\cdots\right],
\end{equation}
\begin{equation}
\label{eq:pr2}
\frac{1}{(l-p_{1,2})^2-m_{1,2}^2}=\frac{1}{r_{2,1}s^2}\left[1-\tilde B_S\frac{(2r_{1,2}-1)}{r_{2,1}}-
\frac{2r_{1,2}M^2}{r_{2,1}s^2}(\tilde B_{AV}-
r_{1,2}\tilde B_S)-
\frac{(p^2\mp 2lp)}{r_{2,1}s^2}+\cdots\right],
\end{equation}
\begin{equation}
\label{eq:pr3}
\frac{1}{k_{1,2}^2}=\frac{1}{r_{2,1}^2s^2}\left[1-\frac{(1-2r_{2,1})}{r_{2,1}}(\tilde B_S+
\tilde B_{AV})\pm
\frac{2(pQ+qP)}{r_{2,1}s^2}-\frac{(p^2+q^2+2pq)}{r^2_{2,1}s^2}+\cdots\right],
\end{equation}
where $B_S$ and $B_{AV}$ are the bound state energies of scalar and vector diquarks,
$\tilde B_{S,AV}=B_{S,AV}/(m_1+m_2)$,
$M=m_1+m_2$, $r_{1,2}=m_{1,2}/M$.
Substituting \eqref{eq:pr1}-\eqref{eq:pr3}, \eqref{eq:amp1}-\eqref{eq:amp2} in~\eqref{eq:amp} we preserve relativistic
factors entering the denominators of relativistic wave functions~\eqref{eq:amp1} and \eqref{eq:amp2},
but in the numerator of the amplitude~\eqref{eq:amp} we take into
account corrections of second order in $|{\bf p}|/m_{1,2}$ and $|{\bf q}|/m_{1,2}$
relative to the leading order result. This provides the convergence of
resulting momentum integrals. Calculating the trace in the amplitude~\eqref{eq:amp} by means of
the system FORM \cite{FORM}, we find that relativistic amplitudes describing the production of
diquark pairs have the following structure:
\begin{equation}
\label{eq:amp11}
{\cal M}_{SS}=-\frac{128\pi^2\alpha}{3s^6}\frac{M^5}{r_1^2r_2^2M_{D_{bc}}^3}(v_2-v_1)^\beta\bar v(p_+)
\gamma_\beta u(p_-)\delta_{ij}
\bar\Psi^0_{SD_{bc}}(0)\bar\Psi^0_{S\bar D_{\bar b\bar c}}(0)\times
\end{equation}
\begin{displaymath}
\left[\frac{Q_c\alpha_s(\frac{m_2^2}{M^2}s^2)}{r_2^3}F_{1,S}+
\frac{Q_b\alpha_s(\frac{m_1^2}{M^2}s^2)}{r_1^3}F_{2,S}
\right]
\end{displaymath}
\begin{equation}
\label{eq:amp22}
{\cal M}_{SAV}=-\frac{128\pi^2\alpha}{3s^6}\frac{M^5}{M_{D_{bc}}^{3/2}M_{\bar D_{\bar b\bar c}}^{3/2}}
\varepsilon_{\beta\alpha\sigma\lambda}
\varepsilon_{\cal AV}^\alpha v_1^\sigma
v_2^\lambda\bar v(p_+)\gamma_\beta u(p_-)\delta_{ij}
\bar\Psi^0_{SD_{bc}}(0)\bar\Psi^0_{AV\bar D_{\bar b\bar c}}(0)\times
\end{equation}
\begin{displaymath}
\left[\frac{Q_c\alpha_s(\frac{m_2^2}{M^2}s^2)}{r_2^3}F_{1,SAV}-
\frac{Q_b\alpha_s(\frac{m_1^2}{M^2}s^2)}{r_1^3}F_{2,SAV}
\right],
\end{displaymath}
\begin{displaymath}
{\cal M}_{AVAV}=-\frac{128\pi^2\alpha}{3s^6}\frac{M^5}{r_1^2r_2^2M_{D_{bc}}^3}
\bar v(p_+)\gamma_\beta u(p_-)\delta_{ij}
\bar\Psi^0_{AVD_{bc}}(0)\bar\Psi^0_{AV\bar D_{\bar b\bar c}}(0)
\Bigl[F_{1,AV}(v_2-v_1)^\beta(\varepsilon_{1,{\cal AV}}\cdot\varepsilon_{2,{\cal AV}})+
\end{displaymath}
\begin{equation}
\label{eq:amp33}
+F_{2,AV}(v_2-v_1)^\beta(\varepsilon_{1,{\cal AV}}\cdot v_2)(\varepsilon_{2,{\cal AV}}\cdot v_1)
+F_{3,AV}\left[(\varepsilon_{2,{\cal AV}}\cdot v_1)\varepsilon^{\beta}_{1,{\cal AV}}-
(\varepsilon_{1,{\cal AV}}\cdot v_2)\varepsilon^{\beta}_{2,{\cal AV}}\right]\Bigr],
\end{equation}
where $\varepsilon_{1,2,{\cal AV}}$ are the polarization vectors of spin 1 diquarks.
The coefficient functions $F_{i,S}$, $F_{i,SAV}$, $F_{i,AV}$  can be presented
as sums of terms containing specific relativistic factors $C_{ij}=[(m_1-\epsilon_1(p))/(m_1+\epsilon_1(p))]^i
[(m_2-\epsilon_2(q))/(m_2+\epsilon_2(q))]^j$ with $i+j\leq 2$.
Used analytical expressions for these functions are written explicitly in Appendix A.
Introducing the scattering angle $\theta$ between the electron momentum ${\bf p}_e$ and momentum ${\bf P}$
of diquark $D_{bc}$, we can calculate the differential cross section
$d\sigma/d\cos\theta$ and then the total cross section $\sigma$ as a
function of center-of-mass energy $s$, masses of quarks and diquarks and relativistic
parameters presented below. We find it useful to write double heavy diquark production differential
cross sections in the following form:
\begin{equation}
\label{eq:sech1}
\frac{d\sigma_{SS}}{d\cos\theta}=\frac{256\pi^3\alpha^2}{3s^{10}}\frac{M^{8}}{M_{D_{bc}}^6r_1^4r_2^4}
|\bar\Psi^0_{SD_{bc}}(0)|^2|\bar\Psi_{{S\bar D}_{\bar b\bar c}}(0)|^2\left(1-
\frac{4M^2_{D_{bc}}}{s^2}\right)^{3/2}\left(1-\cos^2\theta\right)\times
\end{equation}
\begin{displaymath}
\left[\frac{Q_c\alpha_s(\frac{m_2^2}{M^2}s^2)}{r_2^3}F_{1,S}+
\frac{Q_b\alpha_s(\frac{m_1^2}{M^2}s^2)}{r_1^3}F_{2,S}
\right]^2,
\end{displaymath}
\begin{equation}
\label{eq:sech2}
\frac{d\sigma_{SAV}}{d\cos\theta}=\frac{64\pi^3\alpha^2}{3s^8}\frac{M^6}{M_{D_{bc}}^3M_{\bar D_{\bar b\bar c}}^3}
\left[\left(1-\frac{(M_{D_{bc}}+M_{\bar D_{\bar b\bar c}})^2}{s^2}\right)\left(1-\frac{(M_{D_{bc}}-
M_{\bar D_{\bar b\bar c}})^2}{s^2}\right)\right]^{3/2}\times
\end{equation}
\begin{displaymath}
|\bar\Psi^0_{SD_{bc}}(0)|^2|\bar\Psi_{{AV\bar D}_{\bar b\bar c}}(0)|^2
\left[\frac{Q_c\alpha_s\left(\frac{m_2^2}{M^2}s^2\right)}{r_2^3}F_{1,SAV}-
\frac{Q_b\alpha_s\left(\frac{m_1^2}{M^2}s^2\right)}{r_1^3}F_{2,SAV}\right]^2
\left(2-\sin^2\theta\right),
\end{displaymath}
\begin{equation}
\label{eq:sech3}
\frac{d\sigma_{AVAV}}{d\cos\theta}=\frac{64\pi^3\alpha^2}{3s^{10}}\frac{M^8}{M_{D_{bc}}^6r_1^4r_2^4}
|\bar\Psi^0_{AVD_{bc}}(0)|^4\left(1-\frac{4M^2_{D_{bc}}}{s^2}\right)^{3/2}\left(F_A-F_B\cdot\cos^2\theta\right),
\end{equation}
\begin{displaymath}
F_A=F^2_{1,AV}(12-4\eta+\eta^2)+F_{1,AV}F_{2,AV}(8\eta-6\eta^2+\eta^3)+F_{1,AV}F_{3,AV}(4\eta-2\eta^2)+
\end{displaymath}
\begin{displaymath}
+F^2_{2,AV}(4\eta^2-2\eta^3+\frac{1}{4}\eta^4)+F_{2,AV}F_{3,AV}(4\eta^2-\eta^3)+F^2_{3,AV}(2\eta+\eta^2),
\end{displaymath}
\begin{displaymath}
F_B=F^2_{1,AV}(12-4\eta+\eta^2)+F_{1,AV}F_{2,AV}(8\eta-6\eta^2+\eta^3)+F_{1,AV}F_{3,AV}(4\eta-2\eta^2)+
\end{displaymath}
\begin{displaymath}
+F^2_{2,AV}(4\eta^2-2\eta^3+\frac{1}{4}\eta^4)+F_{2,AV}F_{3,AV}(4\eta^2-\eta^3)+F^2_{3,AV}(-2\eta+\eta^2),
\end{displaymath}
where $\eta=s^2/M^2_{D_{bc}}$, the values of wave function at the origin are equal
\begin{equation}
\Psi^0_{S,AVD_{bc}}(0)=\int \sqrt{\frac{(\epsilon_1(p)+m_1)(\epsilon_2(p)+m_2)}
{2\epsilon_1(p)\cdot 2\epsilon_2(p)}}\Psi^0_{S,AVD_{bc}}({\bf p})\frac{d{\bf p}}{(2\pi)^3}.
\end{equation}

This form of differential cross sections is very close to nonrelativistic form obtained in \cite{bkc}.
In nonrelativistic limit our results coincide with the calculations made in \cite{bkc} excepting the cross
section~\eqref{eq:sech3}, which differs by the factor $1/8$ from \cite{bkc}\footnote{We are grateful
to V.V.~Braguta
for the discussion of results obtained in \cite{bkc}}.
The functions $F_{i,S}$, $F_{i,SAV}$ and $F_{i,AV}$ are obtained as series in $|{\bf p}|/m_{1,2}$ and
$|{\bf q}|/m_{1,2}$
up to corrections of second order.
Relativistic parameters $\omega^{S,AV}_{nk}$ entering in $F_{i,S}$, $F_{i,SAV}$ and $F_{i,AV}$
(see Appendix A)
can be expressed in terms of momentum integrals $I_{nk}$ as follows:
\begin{equation}
\label{eq:intnk}
I_{nk}^{S,AV}=\int_0^\infty q^2R^{S,AV}_{D_{bc}}(q)\sqrt{\frac{(\epsilon_1(q)+m_1)(\epsilon_2(q)+m_2)}
{2\epsilon_1(q)\cdot 2\epsilon_2(q)}}
\left(\frac{m_1-\epsilon_1(q)}{m_1+\epsilon_1(q)}\right)^n
\left(\frac{m_2-\epsilon_2(q)}{m_2+\epsilon_2(q)}\right)^k dq,
\end{equation}
\begin{equation}
\label{eq:parameter}
\omega^{S,AV}_{10}=\frac{I^{S,AV}_{10}}{I^{S,AV}_{00}},~\omega^{S,AV}_{01}=\frac{I^{S,AV}_{01}}{I^{S,AV}_{00}},
~\omega^{S,AV}_{11}=\frac{I^{S,AV}_{11}}{I^{S,AV}_{00}},~
\omega^{S,AV}_{20}=\frac{I^{S,AV}_{20}}{I^{S,AV}_{00}},~\omega^{S,AV}_{02}=\frac{I^{S,AV}_{02}}{I^{S,AV}_{00}}.
\end{equation}

On the one hand, in the potential quark model relativistic corrections, connected with relative motion of
heavy quarks,
enter the production amplitude~\eqref{eq:amp} and the cross sections \eqref{eq:sech1}, \eqref{eq:sech2} and
\eqref{eq:sech3}
through the different relativistic factors. They are determined in
final expressions by specific parameters $\omega^{S,AV}_{nk}$. The
momentum integrals which determine the parameters $\omega^{S,AV}_{nk}$ are convergent and we can calculate
them numerically, using the wave functions obtained by the numerical solution of the Schr\"odinger
equation. Nevertheless, we introduce new cutoff parameter $\Lambda\approx m_c$
for momentum integrals $I_{nk}$ in~\eqref{eq:intnk} at high momenta $q$ because we don't know exactly
the bound state wave functions in the region of the relativistic momenta.

The exact form of the wave functions $\Psi^0_{SD_{bc}}({\bf q})$ and $\Psi^0_{AVD_{bc}}({\bf q})$ is important
to improve an accuracy of the calculation of relativistic effects. In nonrelativistic approximation
double diquark production cross sections~\eqref{eq:sech1},  \eqref{eq:sech2} and \eqref{eq:sech3} contain
fourth power of nonrelativistic wave function at the origin.
Small changes of $\Psi^0_{S,AVD_{bc}}$ lead to substantial changes of final results. In the framework of
NRQCD this problem is closely related to the determination of color-singlet matrix elements for
heavy quarkonium \cite{BBL}. Thus, on the other hand, there are
relativistic corrections to the bound state wave functions of scalar and axial vector diquarks. In order to
take them into account, we suppose that the dynamics of a $(bc)$-pair is determined by the QCD generalization of
the standard Breit Hamiltonian in the center-of-mass reference frame \cite{repko1,pot1,pot3,capstick}:
\begin{equation}
\label{eq:breit}
H=H_0+\Delta U_1+\Delta U_2,~~~H_0=\sqrt{{\bf
p}^2+m_1^2}+\sqrt{{\bf
p}^2+m_2^2}-\frac{2\tilde\alpha_s}{3r}+\frac{1}{2}(Ar+B),
\end{equation}
\begin{equation}
\label{eq:breit1}
\Delta U_1(r)=-\frac{\alpha_s^2}{6\pi r}\left[2\beta_0\ln(\mu
r)+a_1+2\gamma_E\beta_0
\right],~~a_1=\frac{31}{3}-\frac{10}{9}n_f,~~\beta_0=11-\frac{2}{3}n_f,
\end{equation}
\begin{equation}
\label{eq:breit2}
\Delta U_2(r)=-\frac{\alpha_s}{3m_1m_2r}\left[{\bf p}^2+\frac{{\bf
r}({\bf r}{\bf p}){\bf p}}{r^2}\right]+\frac{\pi
\alpha_s}{3}\left(\frac{1}{m_1^2}+\frac{1}{m_2^2}\right)\delta({\bf r})+
\frac{2\alpha_s}{3r^3}\left(\frac{1}{2m_1^2}+\frac{1}{m_1m_2}\right)({\bf S}_1{\bf L})+
\end{equation}
\begin{displaymath}
+\frac{2\alpha_s}{3r^3}\left(\frac{1}{2m_2^2}+\frac{1}{m_1m_2}\right)({\bf S}_2{\bf L})
+\frac{16\pi\alpha_s}{9m_1m_2}({\bf S}_1{\bf S}_2)\delta({\bf r})+
\frac{2\alpha_s}{m_1m_2r^3}\left[\frac{({\bf S}_1{\bf r})({\bf S}_2{\bf r})}{r^2}-
\frac{1}{3}({\bf S}_1{\bf S}_2)\right]-
\end{displaymath}
\begin{displaymath}
-\frac{\alpha_s^2(m_1+m_2)}{2m_1m_2r^2}\left[1-\frac{4m_1m_2}{9(m_1+m_2)^2}\right],
\end{displaymath}
where ${\bf L}=[{\bf r}\times{\bf p}]$, ${\bf S}_1$, ${\bf S}_2$ are spins of heavy quarks,
$n_f$ is the number of flavors, $\gamma_E\approx 0.577216$ is
the Euler constant. To describe the hyperfine splittings in $(b\bar c)$ and $(c\bar c)$ mesons
(and $S$-wave diquark system)
which could be in agreement with experimental data and other calculations in quark models
we add to the standard Breit potential
the spin confining potentials obtained in \cite{repko1,repko2}:
\begin{equation}
\Delta V^{hfs}_{conf}(r)=
f_V\frac{A}{8r}\left\{\frac{1}{m_1^2}+\frac{1}{m_2^2}+\frac{16}{3m_1m_2}({\bf S}_1{\bf S}_2)+
\frac{4}{3m_1m_2}\left[3({\bf S}_1 {\bf r}) ({\bf S}_2 {\bf r})-({\bf S}_1 {\bf S}_2)\right]\right\},
\end{equation}
where we take the parameter $f_V=0.9$. For the dependence of the
QCD coupling constant $\tilde\alpha_s(\mu^2)$ on the renormalization point
$\mu^2$ in the pure Coulomb term in~\eqref{eq:breit} we use the three-loop result \cite{kniehl1997}
\begin{equation}
\tilde\alpha_s(\mu^2)=\frac{4\pi}{\beta_0L}-\frac{16\pi b_1\ln L}{(\beta_0 L)^2}+\frac{64\pi}{(\beta_0L)^3}
\left[b_1^2(\ln^2 L-\ln L-1)+b_2\right], \quad L=\ln(\mu^2/\Lambda^2),
\end{equation}
whereas in other terms of the Hamiltonians~\eqref{eq:breit1} and \eqref{eq:breit2} we take
the leading order approximation. The typical momentum transfer scale in a
quarkonium is of order of double reduced mass, so we set the renormalization scale $\mu=2m_1m_2/(m_1+m_2)$ and
$\Lambda=0.168$ GeV, which gives $\alpha_s=0.314$ for diquark $(cc)$, $\alpha_s=0.265$ for diquark $(bc)$.
The coefficients $b_i$ are written explicitly in \cite{kniehl1997}.
The parameters of the linear potential $A=0.18$ GeV$^2$ and
$B=-0.16$ GeV have usual values of quark models.

\begin{table}[h]
\caption{Numerical values of relativistic parameters~\eqref{eq:parameter}
in double heavy diquark production cross sections~\eqref{eq:sech1}, \eqref{eq:sech2}, \eqref{eq:sech3}.}
\bigskip
\label{tb1}
\begin{ruledtabular}
\begin{tabular}{|c|c|c|c|c|c|c|c|c|}
Diquarks&$n^{2S+1}L_J$ &$M_{D}$, &$\Psi^0_{S,AVD}(0)$, & $\omega^{S,AV}_{10}$ &$\omega^{S,AV}_{01}$ &
$\omega^{S,AV}_{11}$  &  $\omega^{S,AV}_{02}$ &  $\omega^{S,AV}_{20}$ \\
$(bc)$, $(cc)$,  &     &   GeV  &   GeV$^{3/2}$  &    &     &    &    &   \\    \hline
$SD_{bc}$&$1^1S_0$ & 6.349 & 0.148 & -0.0454 & -0.0054  & 0.00048  & 0.00006   & 0.0039  \\  \hline
$AVD_{bc}$  & $1^3S_1$  & 6.362 & 0.136 & -0.0467  & -0.0055   &  0.00048  &0.00006   &  0.0039 \\  \hline
$AVD_{cc} $    &$1^3S_1$  & 3.339 & 0.114 & -0.0431  & -0.0431  & 0.0033 & 0.0033  & 0.0033    \\  \hline
\end{tabular}
\end{ruledtabular}
\end{table}

\begin{table}[h]
\caption{The comparison of obtained results for the production cross sections with nonrelativistic calculation.
In third column we present nonrelativistic result obtained in our model pointing out the Ref.\cite{bkc}
where nonrelativistic approximation of the cross sections was discussed for the first time.}
\bigskip
\label{tb2}
\begin{ruledtabular}
\begin{tabular}{|c|c|c|c|}
Final state $D_1D_2$ & Center-of-mass energy s& $\sigma_{NR}$, \cite{bkc}& Our result $\sigma_R$ \\    \hline
$SD_{bc}+S\bar D_{\bar b\bar c}$ & 15.0 GeV & 0.0009 fb & 0.0011 fb  \\  \hline
$SD_{bc}+AV\bar D_{\bar b\bar c}$ & 16.0 GeV& 0.070 fb &  0.034 fb\\  \hline
$AVD_{bc}+AV\bar D_{\bar b\bar c}$ & 16.0 GeV& 0.178 fb & 0.072  fb \\  \hline
$AVD_{cc}+AV\bar D_{\bar c\bar c}$ & 7.6 GeV &0.378 fb  &  0.095 fb \\  \hline
$AVD_{cc}+AV\bar D_{\bar c\bar c}$ & 10.6 GeV &0.070 fb  &  0.025 fb \\  \hline
\end{tabular}
\end{ruledtabular}
\end{table}

\begin{figure}[t!]
\centering
\includegraphics[width=7.5 cm]{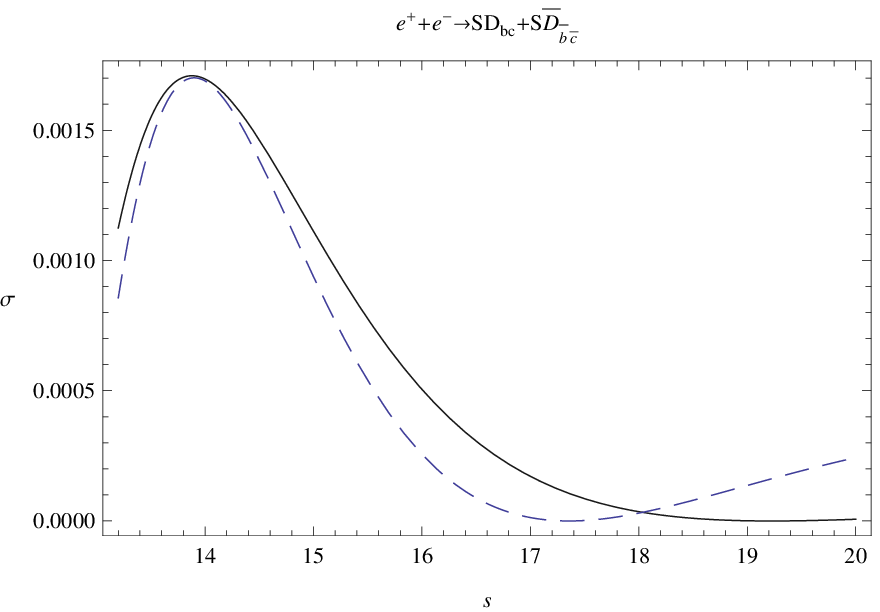}
\includegraphics[width=7.5 cm]{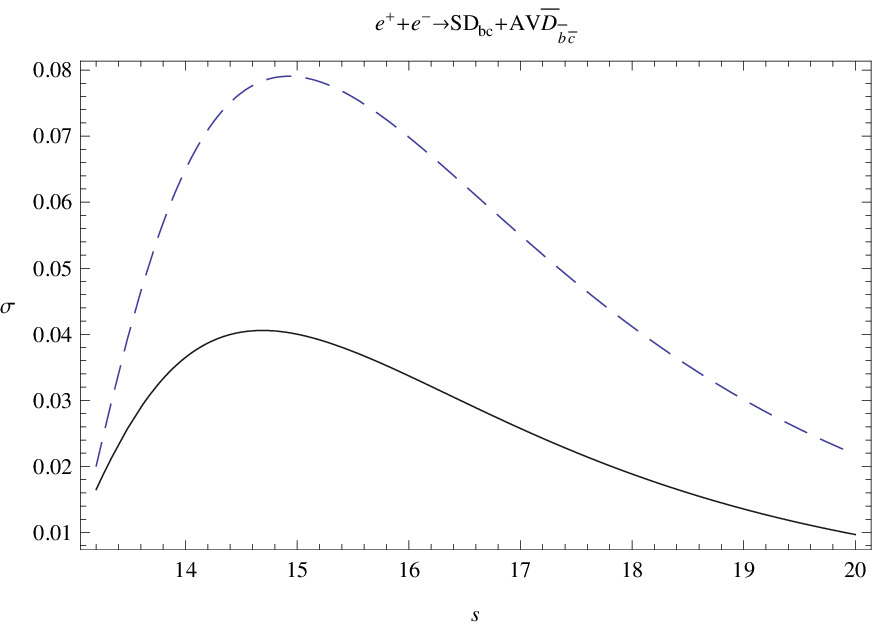}
\includegraphics[width=7.5 cm]{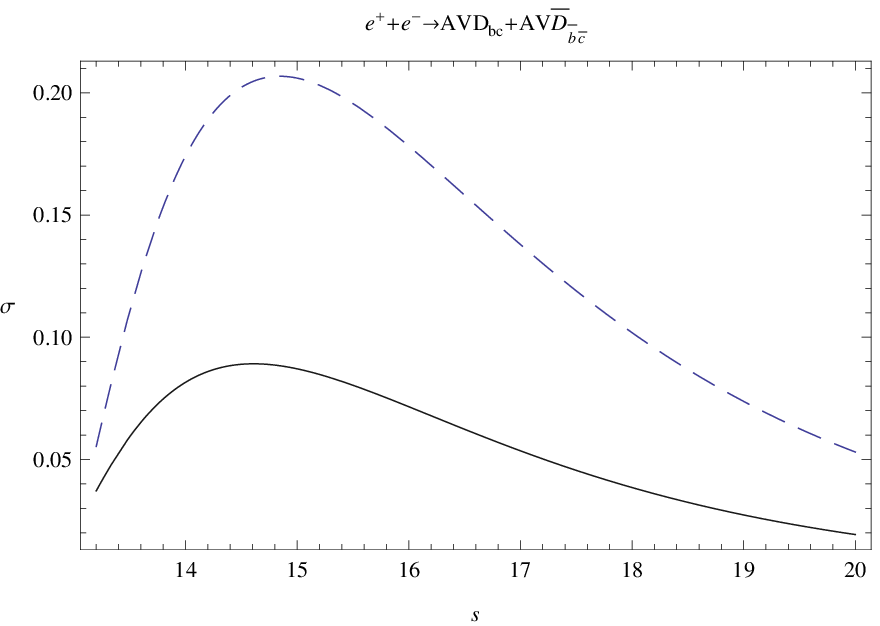}
\includegraphics[width=7.5 cm]{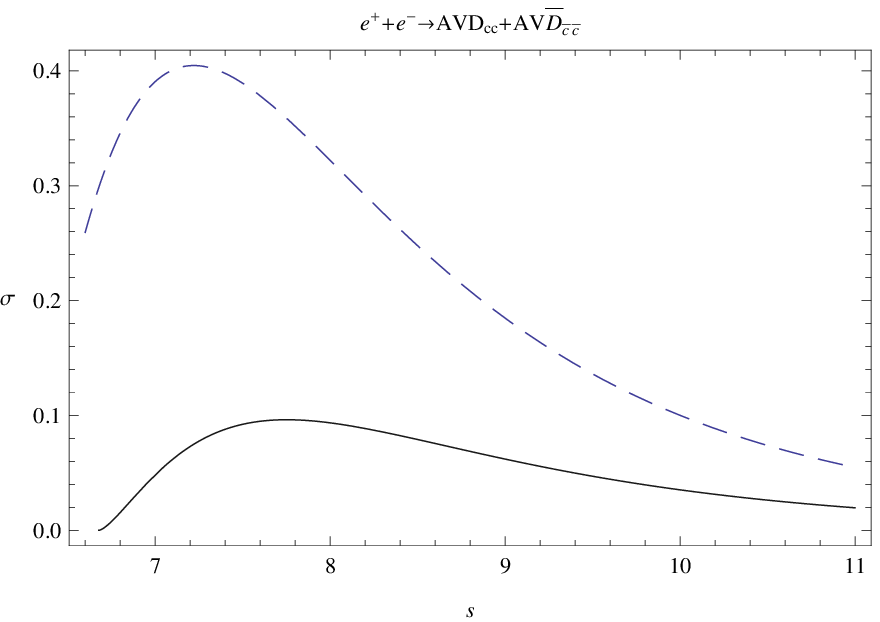}
\caption{The cross section in fb of $e^+e^-$ annihilation into a pair
of $S$-wave scalar and axial vector diquark states $(bc)$ and
axial vector diquark state $(cc)$
as a function of the center-of-mass energy
$s$ (solid line). The dashed line shows nonrelativistic result without
bound state and relativistic corrections.}
\label{fig:fig2}
\end{figure}

For the calculation of relativistic corrections in the bound state diquark wave functions
$\Psi^0_{S,AVD_{bc}}({\bf p})$ we take the Breit potential~\eqref{eq:breit} and
construct the effective potential model as in~\cite{EM2010,Lucha} by means of
the rationalization of kinetic energy operator.
Using the program of numerical solution of the Schr\"odinger equation~\cite{LS}
we obtain the values of all
relativistic parameters entering the cross sections~\eqref{eq:sech1}, \eqref{eq:sech2} and \eqref{eq:sech3}
which are collected in Table~\ref{tb1}.
There is no free diquark to study the effective interaction between two heavy quarks.
So, as a test calculation for our model we find the masses of charmonium states and $B_c$ mesons which are in good
agreement with experimental data and other calculations in quark models. For example, in the case of low lying
$(b\bar c)$ mesons
we obtain $M(B_c^\pm)=M(1^1S_0)=6.280$ GeV and $M(1^3S_1)=6.322$ GeV. Numerical data related with
charmonium states are discussed in~\cite{EM2010}. Strictly speaking we can obtain the charmonium mass
spectrum which agrees with experimental data with more than a per cent accuracy \cite{EM2010,PDG}. Our masses of $S$-wave diquarks
$(bc)$ and $(cc)$ in nonrelativistic approximation are 6.608 GeV and 3.328 GeV correspondingly. In \cite{gklo}
a diquark $(bc)$ ($1S$-state) has the mass 6.48 GeV and diquark $(cc)$ ($1S$-state) 3.16 GeV. The difference
between \cite{gklo} and our results amounts near 2 and 4 per cents and is related with the different value of $c$-quark
mass in \cite{gklo}.
An account of relativistic corrections in our model leads to
slightly different values: the mass of $(bc)$ diquark is 6.349 GeV ($S$=0-state), 6.362 GeV ($S$=1-state)
and mass of $(cc)$ diquark is 3.339 GeV ($S$=1-state).
The difference in $3\div 4$ per cents occurs in comparison with our results in \cite{baryon_diquark} where different
approach to the calculation of relativistic corrections is used.
The values of diquark $(bc)$ and $(cc)$ wave functions at the origin in \cite {gklo} $\Psi_{bc}(0)=0.205~GeV^{3/2}$
and $\Psi_{cc}(0)=0.150~GeV^{3/2}$ are in the agreement with our nonrelativistic results $\Psi^0_{D_{bc}}(0)=0.185~GeV^{3/2}$
and $\Psi^0_{D_{cc}}(0)=0.145~GeV^{3/2}$.
Then we calculate the parameters of diquark states and production cross sections as functions of
center-of-mass energy $s$.
The total cross section plots for the production of diquarks $(bc)$ and $(cc)$ are presented in Fig.~\ref{fig:fig2}.
In Table~\ref{tb2} we give numerical values of total production cross sections at certain center-of-mass
energies $s$ and
compare them with nonrelativistic result in our quark model. These numerical results could be considered
as an estimate for the experimental search.

\section{Numerical results and discussion}

In this paper we have investigated the role of relativistic and bound state effects
in the production processes of a pair double heavy diquarks in
the quark model. We calculate relativistic effects taking into account their
important role in the exclusive pair production of charmonium states in $e^+e^-$ annihilation.
By the construction of the production amplitude~\eqref{eq:amp}
we keep relativistic corrections of two types. The first type is
determined by several functions depending on the relative quark
momenta  ${\bf p}$ and ${\bf q}$ arising from the gluon propagator,
the quark propagator and the relativistic diquark wave functions. The
second type of corrections originates from the perturbative and nonperturbative
treatment of the quark-quark interaction operator which leads to
the different wave functions $\Psi^0_{S,AVD_{bc}}({\bf p})$ and
$\Psi^0_{S,AV\bar D_{\bar b\bar c}}({\bf q})$ for the diquark bound states.
In addition, we systematically accounted for
the bound state corrections working with the masses of diquark bound states
or with the bound state energies $B_S$, $B_{AV}$.
The calculated masses of diquark states agree well with previous theoretical results \cite{baryon_diquark}.
Note that basic parameters of the model are kept fixed from
previous calculations of the meson mass spectra and decay widths
\cite{rqm5,rqm1,brambilla-2011,QWG}.

It follows from the results~\eqref{eq:sech1}, \eqref{eq:sech2}, \eqref{eq:sech3} that the
total cross sections for the exclusive pair production of scalar, scalar+axial vector, axial
vector diquarks in $e^+e^-$ annihilation can be presented in the form:
\begin{equation}
\label{eq:sech1t}
\sigma_{SS}=\frac{1024\pi^3\alpha^2}{9s^{10}}\frac{M^{8}}{r_1^4r_2^4M^6_{D_{bc}}}|\bar\Psi^0_{SD_{bc}}(0)|^4
\left(1-\frac{4M^2_{D_{bc}}}{s^2}\right)^{3/2}\left[\frac{Q_c\alpha_s(\frac{m_2^2}{M^2}s^2)}{r_2^3}F_{1,S}+
\frac{Q_b\alpha_s(\frac{m_1^2}{M^2}s^2)}{r_1^3}F_{2,S}
\right]^2,
\end{equation}
\begin{equation}
\label{eq:sech2t}
\sigma_{SAV}=\frac{512\pi^3\alpha^2}{9s^8}\frac{M^{6}}{M_{D_{bc}}^3M_{\bar D_{\bar b\bar c}}^3}
\left[\left(1-\frac{(M_{D_{bc}}+M_{\bar D_{\bar b\bar c}})^2}{s^2}\right)\left(1-\frac{(M_{D_{bc}}-
M_{\bar D_{\bar b\bar c}})^2}{s^2}\right)\right]^{3/2}\times
\end{equation}
\begin{displaymath}
|\bar\Psi^0_{SD_{bc}}(0)|^2|\bar\Psi_{{AV\bar D}_{\bar b\bar c}}(0)|^2
\left[\frac{Q_c\alpha_s\left(\frac{m_2^2}{M^2}s^2\right)}{r_2^3}F_{1,SAV}-
\frac{Q_b\alpha_s\left(\frac{m_1^2}{M^2}s^2\right)}{r_1^3}F_{2,SAV}\right]^2,
\end{displaymath}
\begin{equation}
\label{eq:sech3t}
\sigma_{AVAV}=\frac{128\pi^3\alpha^2}{9s^{10}}\frac{M^8}{r_1^4r_2^4M_{D_{bc}}^6}
|\bar\Psi^0_{AVD_{bc}}(0)|^4\left(1-\frac{4M^2_{D_{bc}}}{s^2}\right)^{3/2}\left(3F_A-F_B\right).
\end{equation}

Relativistic corrections to the bound state wave functions, relativistic corrections to the
production amplitudes, bound state effects impact differently on the value of cross sections.
In Fig.~\ref{fig:fig2} we show the plots of total cross sections corresponding to the pairs of diquarks scalar+scalar,
scalar+axial vector, axial vector+axial vector as functions of center-of-mass energy $s$. Some kind of
experimental data regarding to such reactions are absent at present, so, these plots could serve only for the estimate
of possible value of cross sections. Among discussed reactions there are maximal numerical values of
the cross section in the case of a pair axial vector diquarks $(bc)$ and $(cc)$ production (this result qualitatively
agrees with that one obtained in \cite{bkc}). So, this production process
could be interested for us first of all because it can have the experimental perspective. Assuming that the luminosity
at the B-factory
${\mathcal L}=10^{34}~cm^{-2}\cdot c^{-1}$ the yield of pairs of double heavy baryons $(ccq)$ can be near $30$ events per year
at the center-of-mass energy $s=7.6~GeV$.
This value is more then by an order of magnitude smaller then that given in \cite{bkc}.
As is mentioned in previous section
the main difference is related with a factor $1/8$. Moreover, an accounting of relativistic and bound state corrections leads
to additional decrease compared with nonrelativistic result. It is necessary to point out that we call nonrelativistic
result that one which is obtained with pure nonrelativistic Hamiltonian when the bound state mass is taken to be $M_{D_{bc}}=m_1+m_2$.
Essential decrease of relativistic cross section value in the case of pair axial diquarks production compared with nonrelativistic
result (see Table II) complicates an observation of such events.
There are several important factors which influence
strongly on the total result when passing from nonrelativistic to relativistic theory. Relativistic corrections to the
production amplitude increase nonrelativistic result. This is true for all cross sections. But another relativistic corrections
to the bound state wave functions and bound state corrections have an opposite effect. They lead to essential decrease of the wave function at the origin
and, as a result, to decrease of the production cross sections in the case of SD+AVD and AVD+AVD cross sections.
So, for $(cc)$-diquark an account of spin-dependent relativistic corrections leads to a decrease of the cross section by a factor
$\Psi^0_{D_{cc},ner}(0)/\Psi^0_{D_{cc}}(0)\approx 2.7$. A diquark is a more bulky object as compared with a meson so,
decreasing factors become significantly stronger than in the meson case. Note that in the case of the production of a diquark with two
identical quarks it is necessary to take into account the Pauli exclusion principle. This means that
we should introduce in the production amplitude additional factor $1/2$ for each pair $(cc)$ and $(\bar c\bar c)$.

Making the estimate of a pair of baryons production we suppose that a spin-1 diquark $(cc)$ can fragment either to a spin $J=1/2$ baryon $(ccq)$
containing light quark $u$, $d$,
which we denote $\Xi_{cc}$ or to a spin $J=3/2$ baryon $(ccq)$ which we denote $\Xi_{cc}^\ast$ baryon.
The production cross section for a pair baryon-antibaryon $(B\bar B)$ is
\begin{equation}
d\sigma_{B\bar B}=\int_0^1 dz_1\int_0^1 dz_2\frac{d\sigma}{dz_1 dz_2}(e^+e^-\to D\bar D)\cdot D_{D\to B}(z_1)
\cdot D_{\bar D\to\bar B}(z_2),
\end{equation}
where $z_i$ is the part of baryon momenta carried out by the diquark. The baryon has approximately the same
momentum as a diquark,
so we can present the diquark fragmentation function $D_{D\to B}(z)$ as follows \cite{falk}:
\begin{equation}
D_{D\to B}(z)=P_{D\to B}\cdot \delta(1-z),
\end{equation}
where $P_{D\to B}$ is the total fragmentation probability of a diquark to a baryon. This probability can be taken
equal to unity for the diquark fragmentation to the baryon $(ccq)$: $\int_0^1 D_{D\to B}(z)dz=1$.
So, obtained above cross sections~\eqref{eq:sech1t},
\eqref{eq:sech2t},
\eqref{eq:sech3t} can be used also for the estimate of a pair baryon-antibaryon production in $e^+e^-$ annihilation.
It is important to note that at high energy $e^+e^-$ colliders the rate for the production of a pair of double heavy
baryon $(ccq)$ -antibaryon $(\bar c\bar c\bar q)$ is comparable with the production rates for $S$ and $P$-wave charmonium states
some of which were observed experimentally.

We presented a treatment of relativistic effects in the
$S$-wave double diquark production in $e^+e^-$ annihilation. Two different types of
relativistic contributions to the production amplitudes~\eqref{eq:amp11}, \eqref{eq:amp22}, \eqref{eq:amp33}
are singled out. The first type includes relativistic $v/c$ corrections to the wave functions and their
relativistic transformations. The second type includes relativistic $p/s$ corrections appearing from the
expansion of the quark and gluon propagators. The latter corrections are taken into account up to the second order.
It is important to note that the expansion parameter $p/s$ is very small. In our analysis
of the production amplitudes we correctly take into account
relativistic contributions of order $O(v^2/c^2)$ for the $S$-wave diquarks. Therefore the first basic
theoretical uncertainty of our calculation is connected with omitted terms of order $O({\bf p}^4/m^4)$.
Since the calculation of masses of $S$-wave diquark states is sufficiently accurate in our
model (a comparison with the meson masses is performed), we suppose that the uncertainty in the cross section
calculation due to omitted relativistic corrections of order $O({\bf p}^4/m^4)$ in the
quark interaction operator (the Breit Hamiltonian) is also very small.
Taking into account that the average value of the heavy quark velocity squared in the
charmonium is $\langle v^2\rangle=0.3$, we expect that relativistic corrections
of order $O({\bf p}^4/m^4)$ to the cross sections~\eqref{eq:sech1t}, \eqref{eq:sech2t}, \eqref{eq:sech3t},
coming from the production amplitude should not exceed $30\%$ of the obtained
relativistic result. Strictly speaking in the quasipotential approach
we can not find precisely the bound state wave functions in the region of relativistic momenta $p\ge m$.
Using indirect arguments related with the mass spectrum calculation we estimate in $10\%$ the uncertainty in the wave function
determination. Larger value of the error will lead to the essential discrepancy between the experiment and theory in the
calculation of the charmonium mass
spectrum. Then the corresponding error in the cross sections~~\eqref{eq:sech1t}, \eqref{eq:sech2t}, \eqref{eq:sech3t} is
not exceeding $40\%$.
Another important part of total theoretical error is related with
radiative corrections of order $\alpha_s$ which were omitted in our analysis.
Our approach to the calculation of the amplitude of double diquark production
can be extended beyond the leading order in the strong coupling constant. Then the vertex
functions in~\eqref{eq:amp} will have more complicate structure including the integration over the
loop momenta. Our calculation of the cross
sections accounts for effectively only some part of one loop corrections by means of the
Breit Hamiltonian. So, we assume that radiative corrections of order $O(\alpha_s)$
can cause the $20\%$ modification of the production cross sections.
We have neglected terms in
the cross sections~\eqref{eq:sech1t}, \eqref{eq:sech2t}, \eqref{eq:sech3t} containing the product of $I_{nk}$
with summary index $>2$ because their contribution has been found negligibly small. There are no another
comparable uncertainties related to other parameters of the model, since their values were fixed from our
previous consideration of meson and baryon properties \cite{rqm1,rqm5}. Our total maximum theoretical errors are
estimated in $54\%$. To obtain this estimate we add the above mentioned uncertainties in quadrature.

\acknowledgments

The authors are grateful to V.V.~Braguta, D.~Ebert, R.N.~Faustov and V.O.~Galkin
for useful discussions. The work is performed under the
financial support of the Ministry of Education and Science of Russian Federation
(government order for Samara State U. grant No. 2.870.2011).

\appendix

\section{The coefficient functions $F_{i,S}$, $F_{i,SAV}$ and $F_{i,AV}$ entering in
the production amplitudes (14)-(16)}

General structure of the pair double heavy diquark production amplitudes studied in
this work is the following:
\bga
\label{eq:A1}
\mathcal M=-\frac{8\pi^2\alpha}{3s}\sqrt{M_{D_{bc}}M_{\bar D_{\bar b\bar c}}}\,[\bar v(p_+)
\gamma_\beta u(p_-)]\delta^{ij}\times \\
\int\!\frac{d\mathbf{p}}{(2\pi)^3}\int\!\frac{d\mathbf{q}}{(2\pi)^3}
\frac{\bar\Psi^0_{D_{bc}}({\bf p})}
{\sqrt{\frac{\epsilon_1(p)}{m_1}\frac{(\epsilon_1(p)+m_1)}{2m_1}\frac{\epsilon_2(p)}{m_2}
\frac{(\epsilon_2(p)+m_2)}{2m_2}}}
\frac{\bar\Psi^0_{\bar D_{\bar b\bar c}}({\bf q})}
{\sqrt{\frac{\epsilon_1(q)}{m_1}\frac{(\epsilon_1(q)+m_1)}{2m_1}\frac{\epsilon_2(q)}{m_2}
\frac{(\epsilon_2(q)+m_2)}{2m_2}}}\times \\
\mathrm{Tr}\bigl\{\mathcal T_{12}^\beta+\kappa\,\mathcal T_{34}^\beta\bigr\},
\ega

\bga
\mathcal T_{12}^\beta=\mathcal Q_c\alpha_b\Bigl[\frac{\hat v_1-1}2+\hat v_1\frac{\mathbf p^2}
{2m_2(\epsilon_2(p)+m_2)}-\frac{\hat p}{2m_2}\Bigr]\Sigma^1_{S,AV}(1+\hat v_1)\times\\
\Bigl[\frac{\hat v_1+1}2+\hat v_1\frac{\mathbf p^2}{2m_1(\epsilon_1(p)+m_1)}+\frac{\hat p}{2m_1}
\Bigr]\left[\gamma^\beta\frac{\hat p_1-\hat l+m_1}{(l-p_1)^2-m_1^2}\,\gamma_\mu+\gamma_\mu\,\frac{\hat l-\hat q_1+m_1}{(l-q_1)^2-m_1^2}\gamma^\beta\right]D^{\mu\nu}(k_2)\times \\
\Bigl[\frac{\hat v_2-1}2+\hat v_2\frac{\mathbf q^2}{2m_1(\epsilon_1(q)+m_1)}+\frac{\hat q}{2m_1}
\Bigr]\Sigma^2_{S,AV}(1+\hat v_2)
\Bigl[\frac{\hat v_2+1}2+\hat v_2\frac{\mathbf q^2}{2m_2(\epsilon_2(q)+m_2)}-\frac{\hat q}{2m_2}\Bigr]\gamma_\nu,
\ega

\bga
\mathcal T_{34}^\beta=\mathcal Q_b\alpha_c\Bigl[\frac{\hat v_1-1}2+\hat v_1
\frac{\mathbf p^2}{2m_1(\epsilon_1(p)+m_1)}+\frac{\hat p}{2m_1}\Bigr]\Sigma^1_{S,AV}(1+\hat v_1)\times\\
\Bigl[\frac{\hat v_1+1}2+\hat v_1\frac{\mathbf p^2}{2m_2(\epsilon_2(p)+m_2)}-\frac{\hat p}{2m_2}\Bigr]
\left[
\gamma^\beta\frac{\hat p_2-\hat l+m_2}{(l-p_2)^2-m_2^2}\,\gamma_\mu+
\gamma_\mu\frac{\hat l-\hat q_2+m_2}{(l-q_2)^2-m_2^2}\,\gamma^\beta
\right]
D^{\mu\nu}(k_1)\times \\
\Bigl[\frac{\hat v_2-1}2+\hat v_2\frac{\mathbf q^2}{2m_2(\epsilon_2(q)+m_2)}-
\frac{\hat q}{2m_2}\Bigr]\Sigma^2_{S,AV}(1+\hat v_2)
\Bigl[\frac{\hat v_2+1}2+\hat v_2\frac{\mathbf q^2}{2m_1(\epsilon_1(q)+m_1)}+
\frac{\hat q}{2m_1}\Bigr]\gamma_\nu,
\ega
where $\Sigma^{1,2}_{S,AV}$ is equal to $\gamma_5$ for $S=0$ diquark and
$\hat\varepsilon_{\mathcal{AV}}$ for $S=1$; $\kappa=1$ for the $S-S$ or
$AV-AV$ diquark pair and $\kappa=-1$ for $S-AV$ diquark pair. Calculating the trace in \eqref{eq:A1}
we obtain amplitudes ${\cal M}_{SS}$, ${\cal M}_{SAV}$ and ${\cal M}_{AVAV}$ presented in
Eqs.~\eqref{eq:amp11}-\eqref{eq:amp33}. Corresponding functions $F_{i,S}$, $F_{i,SAV}$ and $F_{i,AV}$
are written below in the used approximation.

\vspace{5mm}

{\underline {$e^++e^-\to SD_{bc}+S\bar D_{\bar b\bar c}$}.

\begin{equation}
F_{1,S}=F_{1,S}^{(0)}+F_{1,S}^{(1)}\omega^S_{10}+F_{1,S}^{(2)}\omega^S_{11}+F_{1,S}^{(3)}
\omega^S_{20}+F_{1,S}^{(4)}(\omega^S_{10})^2+F_{1,S}^{(5)}\tilde B_S,
\end{equation}
\begin{equation}
F_{1,S}^{(0)}=r_2^2(r_2-1)^3+(r_2-1)^2\frac{r_2^3}{2} \tilde s^2,
\end{equation}
\begin{equation}
F_{1,S}^{(1)}=
\left(\frac{5 \text{$r_2$}^4}{3}-7 \text{$r_2$}^3+11 \text{$r_2$}^2-\frac{23 \text{$r_2$}}{3}+2\right)
r_2\tilde s^2
+\frac{-4 \text{$r_2$}^6+12 \text{$r_2$}^5-40\text{$r_2$}^3+60 \text{$r_2$}^2-36 \text{$r_2$}+8}{\tilde s^2}-
\end{equation}
\begin{displaymath}
-2 \text{$r_2$}^6+\frac{11 \text{$r_2$}^5}{3}+\frac{13 \text{$r_2$}^4}{3}-\frac{43 \text{$r_2$}^3}{3}+\frac{31 \text{$r_2$}^2}{3}
-\frac{4 \text{$r_2$}}{3}-\frac{2}{3},
\end{displaymath}
\begin{equation}
F_{1,S}^{(2)}=\left(\frac{2 \text{$r_2$}^4}{3}-5 \text{$r_2$}^3+10 \text{$r_2$}^2-\frac{23 \text{$r_2$}}{3}+2\right)
r_2\tilde s^2+
\frac{-4 \text{$r_2$}^6+12 \text{$r_2$}^5-40\text{$r_2$}^3+60 \text{$r_2$}^2-36 \text{$r_2$}+8}{\tilde s^2}-
\end{equation}
\begin{displaymath}
-2 \text{$r_2$}^6+\frac{5 \text{$r_2$}^5}{3}+\frac{31 \text{$r_2$}^4}{3}-\frac{61 \text{$r_2$}^3}{3}+\frac{37 \text{$r_2$}^2}{3}
-\frac{4 \text{$r_2$}}{3}-\frac{2}{3},
\end{displaymath}
\begin{equation}
F_{1,S}^{(3)}=-F_{1,S}^{(1)}=F_{1,S}^{(4)},
\end{equation}
\begin{equation}
F_{1,S}^{(5)}=\text{$r_2$}^6+\frac{13 \text{$r_2$}^5}{2}-29 \text{$r_2$}^4+\frac{75 \text{$r_2$}^3}{2}-19 \text{$r_2$}^2+3 \text{$r_2$}+
\end{equation}
\begin{displaymath}
+\left(\frac{7 \text{$r_2$}^3}{2}-8 \text{$r_2$}^2+\frac{11\text{$r_2$}}{2}-1\right)r_2^2 \tilde s^2+
\frac{(2\text{$r_2$}^4-8 \text{$r_2$}^3+12 \text{$r_2$}^2-8 \text{$r_2$}+2)r_2^2}{\tilde s^2},
\end{displaymath}
where $\tilde s=s/M$.
We specially violate the symmetry in quarks $c$ and $b$ making the substitution ${\bf p}^2=
(\epsilon_1(p)-m_1)(\epsilon_1(p)+m_1)$
in order to decrease the size of final expression. The function $F_{2,S}$ can be obtained
from $F_{1,S}$ changing $r_2\leftrightarrow r_1$,
$m_1\leftrightarrow m_2$ and $\omega_{ij}\to\omega_{ji}$.

\vspace{5mm}

{\underline {$e^++e^-\to SD_{bc}+AV\bar D_{\bar b\bar c}$}.
\begin{equation}
F_{1,SAV}=F_{1,SAV}^{(0)}+F_{1,SAV}^{(1)}\omega^S_{10}+F_{1,SAV}^{(2)}\omega^{AV}_{10}+
F_{1,SAV}^{(3)}\omega^S_{01}+F_{1,SAV}^{(4)}\omega^{AV}_{01}+
F_{1,SAV}^{(5)}\omega^S_{20}+F_{1,SAV}^{(6)}\omega^{AV}_{20}+
\end{equation}
\begin{displaymath}
+F_{1,SAV}^{(7)}\omega^{S}_{11}+F_{1,SAV}^{(8)}\omega^{AV}_{11}+
F_{1,SAV}^{(9)}\omega^S_{10}\omega^{AV}_{10}+F_{1,SAV}^{(10)}\tilde B_S+F_{1,SAV}^{(11)}\tilde B_{AV},
\end{displaymath}
\begin{equation}
F_{1,SAV}^{(0)}=1,
\end{equation}
\begin{equation}
F_{1,SAV}^{(1)}=\tilde B_{AV} \left(\frac{r_2-1}{ \tilde s^2}+\frac{2 r_2^2+5 r_2-3}{2r_2}\right)+
\tilde B_{S} \left(\frac{r_2-1}{ \tilde s^2}+\frac{6 r_2-3}{2r_2}\right)+
\end{equation}
\begin{displaymath}
+\frac{5r_2^2-6 r_2+1}{3 r_2^2}-\frac{2 r_2^3-6r_2+4}{r_2^2 \tilde s^2},
\end{displaymath}
\begin{equation}
F_{1,SAV}^{(2)}=\tilde B_{AV}\left(\frac{r_2-1}{\tilde s^2}+\frac{2 r_2^2+5 r_2-3}{2r_2}\right)+
\tilde B_{S} \left(\frac{r_2-1}{ \tilde s^2}+\frac{6 r_2-3}{2r_2}\right)-
\end{equation}
\begin{displaymath}
-\frac{2 r_2^3-6 r_2+4}{ r_2^2 \tilde s^2}-\frac{3 r_2^3-12 r_2^2+10 r_2-1}{3 r_2^2},
\end{displaymath}
\begin{equation}
F_{1,SAV}^{(3)}=\tilde B_{AV}\left(\frac{r_2-1}{ \tilde s^2}+\frac{2 r_2^2+5 r_2-3}{2r_2}
\right)+\tilde B_{S} \left(\frac{r_2-1}{ \tilde s^2}+\frac{6 r_2-3 }{2r_2}\right),
\end{equation}
\begin{equation}
F_{1,SAV}^{(4)}=F_{1,SAV}^{(3)},
\end{equation}
\begin{equation}
F_{1,SAV}^{(5)}=-\frac{5 r_2^2-6 r_2+1 }{3 r_2^2}+\frac{2 r_2^3-6 r_2+4 }{r_2^2 \tilde s^2},
\end{equation}
\begin{equation}
F_{1,SAV}^{(6)}=\frac{2 r_2^3-6 r_2+4}{r_2^2 \tilde s^2}+\frac{3 r_2^3-12 r_2^2+10 r_2-1}{3r_2^2},
\end{equation}
\begin{equation}
F_{1,SAV}^{(7)}=\tilde B_{AV} \left(\frac{r_2-1}{ \tilde s^2}+\frac{3 r_2^2-2 r_2-3}{2r_2}\right)+
\tilde B_{S} \left(\frac{r_2-1}{\tilde s^2}+\frac{2 r_2-3}{2r_2}\right)+
\end{equation}
\begin{displaymath}
+\frac{2 r_2^2-6 r_2+1}{3 r_2^2}-\frac{2 r_2^3-6 r_2+4 }{r_2^2 \tilde s^2},
\end{displaymath}
\begin{equation}
F_{1,SAV}^{(8)}=\tilde B_{AV}\left(\frac{r_2-1}{ \tilde s^2}+\frac{3 r_2^2-2 r_2-3}{2r_2}\right)+
\tilde B_{S} \left(\frac{r_2-1}{\tilde s^2}+\frac{2 r_2-3}{2r_2}\right)+
\end{equation}
\begin{displaymath}
-\frac{2 r_2^3-6 r_2+4}{r_2^2 \tilde s^2}-\frac{3 r_2^3-9 r_2^2+10 r_2-1}{3 r_2^2},
\end{displaymath}
\begin{equation}
F_{1,SAV}^{(9)}=-\frac{(1-r_2)^2(40 r_2^2-80 r_2+27)}{9r_2^4 \tilde s^2}
-\frac{(1-r_2)^2(3 r_2^3-31 r_2^2+34 r_2-4)}{9 r_2^4},
\end{equation}
\begin{equation}
F_{1,SAV}^{(10)}=\frac{r_2-1}{ \tilde s^2}+\frac{10 r_2-3}{2 r_2},
\end{equation}
\begin{equation}
F_{1,SAV}^{(11)}=\frac{r_2-1}{ \tilde s^2}+\frac{r_2^2+12 r_2-3}{2 r_2}.
\end{equation}
In these functions we preserve several terms containing the product of parameters $\omega_{ij}^{S,AV}$ and
bound energies $\tilde B_S$ and $\tilde B_{AV}$ in order to increase the accuracy of the calculation.
Note again that the function $F_{2,AV}$ can be obtained from $F_{1,AV}$ by means of the replacement
$m_1\leftrightarrow m_2$, $r_2\leftrightarrow r_1$ and $\omega_{ij}\to\omega_{ji}$.

\vspace{5mm}

{\underline {$e^++e^-\to AVD_{bc}+AV\bar D_{\bar b\bar c}$}.

\begin{equation}
F_{i,AV}=\left[\frac{Q_c\alpha_s\left(\frac{m_2^2}{M^2}s^2\right)}{r_2^3}F_{i1,AV}+
\frac{Q_b\alpha_s\left(\frac{m_1^2}{M^2}s^2\right)}{r_1^3}F_{i2,AV}\right],~~~i=1,2,3,
\end{equation}
\begin{equation}
F_{11,AV}=F_{11,AV}^{(0)}+F_{11,AV}^{(1)}\omega^{AV}_{10}+F_{11,AV}^{(2)}\omega^{AV}_{11}+
F_{11,AV}^{(3)}\omega^{AV}_{20}+F_{11,AV}^{(4)}(\omega^{AV}_{10})^2+
F_{11,AV}^{(5)}\tilde B_{AV},
\end{equation}
\begin{equation}
F_{11,AV}^{(0)}=(1-r_2)^3 r_2^2,
\end{equation}
\begin{equation}
F_{11,AV}^{(1)}=\frac{1}{3} (1-r_2)^3 \left(7 r_2^2-10 r_2+2\right)-\frac{4 (1-r_2)^3
\left(r_2^3-3 r_2+2\right)}{\tilde s^2},
\end{equation}
\begin{equation}
F_{11,AV}^{(2)}=\frac{1}{3} (1-r_2)^3 \left(r_2^2-10 r_2+2\right)-\frac{4 (1-r_2)^3
\left(r_2^3-3 r_2+2\right)}{\tilde s^2},
\end{equation}
\begin{equation}
F_{11,AV}^{(3)}=-F_{11,AV}^{(1)},
\end{equation}
\begin{equation}
F_{11,AV}^{(4)}=\frac{\left(10 r_2^3-r_2^2-10 r_2+4\right) (1-r_2)^4}{9 r_2^2}+\frac{4
\left(3 r_2^4-17 r_2^3+16
r_2^2+4 r_2-6\right) (1-r_2)^4}{9 r_2^2 \tilde s^2},
\end{equation}
\begin{equation}
F_{11,AV}^{(5)}=-\frac{2 r_2^2(1-r_2)^4}{\tilde s^2}+\frac{1}{2} (1-r_2)^2 r_2
\left(-19 r_2^2+26 r_2-6\right),
\end{equation}
\begin{equation}
F_{21,AV}=F_{21,AV}^{(1)}\omega^{AV}_{10}+F_{21,AV}^{(2)}\omega^{AV}_{11}+F_{21,AV}^{(3)}
\omega^{AV}_{20}+F_{21,AV}^{(4)}(\omega^{AV}_{10})^2,
\end{equation}
\begin{equation}
F_{21,AV}^{(1)}=-\frac{8 (1-r_2)^4 r_2}{\tilde s^2},
\end{equation}
\begin{equation}
F_{21,AV}^{(2)}=F_{21,AV}^{(1)}=-F_{21,AV}^{(3)},
\end{equation}
\begin{equation}
F_{21,AV}^{(4)}=\frac{16 (1-r_2)^4 \left(2 r_2^2-3 r_2+1\right)}{9 r_2 \tilde s^2},
\end{equation}

\begin{equation}
F_{31,AV}=F_{31,AV}^{(0)}+F_{31,AV}^{(1)}\omega^{AV}_{10}+F_{31,AV}^{(2)}\omega^{AV}_{11}+
F_{31,AV}^{(3)}\omega^{AV}_{20}+F_{31,AV}^{(4)}(\omega^{AV}_{10})^2+
F_{31,AV}^{(5)}\tilde B_{AV},
\end{equation}
\begin{equation}
F_{31,AV}^{(0)}=(1-r_2)^2 r_2^2,
\end{equation}
\begin{equation}
F_{31,AV}^{(1)}=\frac{4 \left(r_2^2+r_2-2\right) (1-r_2)^3}{\tilde s^2}+\frac{1}{3}
\left(r_2^2-10 r_2+2\right) (1-r_2)^3,
\end{equation}
\begin{equation}
F_{31,AV}^{(2)}=-\frac{4 \left(r_2^3-3 r_2+2\right) (1-r_2)^2}{\tilde s^2}-\frac{1}{3}
\left(r_2^3-5 r_2^2+12 r_2-2\right) (1-r_2)^2,
\end{equation}
\begin{equation}
F_{31,AV}^{(3)}=-F_{31,AV}^{(1)},
\end{equation}
\begin{equation}
F_{31,AV}^{(4)}=-\frac{8 \left(3 r_2^3-2 r_2^2-6 r_2+3\right) (1-r_2)^4}{9 r_2^2
\tilde s^2}-\frac{\left(5 r_2^3-19 r_2^2+18
r_2-4\right) (1-r_2)^4}{9 r_2^2},
\end{equation}
\begin{equation}
F_{31,AV}^{(5)}=-\frac{2 (1-r_2)^3 r_2^2}{\tilde s^2}-\frac{1}{2} (1-r_2)^2 r_2
\left(r_2^2-22 r_2+6\right).
\end{equation}
Other functions $F_{i2,AV}$ ($i=1,2,3$) can be obtained from $F_{i1,AV}$ using the
replacement $m_1\leftrightarrow m_2$, $r_2\leftrightarrow r_1$ and $\omega_{ij}\to\omega_{ji}$.

\end{document}